\documentclass[%
 aps,
 amsmath,amssymb,
 reprint,%
]{revtex4-2}

\usepackage{graphicx}
\usepackage{dcolumn}
\usepackage{bm}

\usepackage[utf8]{inputenc}
\usepackage[T1]{fontenc}
\usepackage{mathptmx}

\begin{document}
\title{Wide Range Thin-Film Ceramic Metal-Alloy Thermometers with Low Magnetoresistance}

\author{N.A. Fortune}
\email{nfortune@smith.edu}
\author{J. E. Palmer-Fortune}
\email{jfortune@smith.edu}
 \author{A. Trainer}
\affiliation{ 
Department of Physics, Smith College, Northampton MA 01063, USA
}
\author{A. Bangura}
\affiliation{%
National High Magnetic Field Laboratory, Tallahassee, FL 32310, USA
}%
\author{N. Kondedan}
\email{neha.kondedan@fysik.su.se}
\author{A. Rydh}
\affiliation{%
Department of Physics, Stockholm University, AlbaNova University Center,  SE-106 91 Stockholm, Sweden
}
\date{\today}

\begin{abstract}

Many thermal measurements in high magnetic fields --- including heat capacity, thermal conductivity, thermopower, magnetocaloric, and thermal Hall effect measurements ---  require thermometers that are sensitive over a wide temperature range, are low mass, have a rapid thermal response, and have a minimal, easily correctable magnetoresistance.
Here we report the development of a new  granular-metal oxide ceramic composite (cermet) for this purpose  formed  by co-sputtering of the metallic alloy nichrome (Ni$_{0.8}$Cr$_{0.2}$) and the insulator silcon dioxide (SiO$_2$). 
We find that co-sputtering of NiCr alloys with SiO$_2$ in a reactive oxygen + inert argon gas mixture  
produces resistive thin film thermometers sensitive enough to be used in calorimetry and related measurements from room temperature down to below 100\,mK in magnetic fields up to at least 41\,tesla. 

\end{abstract}

\maketitle

\section{INTRODUCTION}
Magnetic-field-dependent small sample calorimetric and related thermal measurements \cite{tagliati_differential_2012, labarre_magnetoquantum_2022} impose a particularly stringent set of requirements on the materials used for resistive thermometry. The thermometers should  (1) have a resistance $R(T)$ that depends monotonically on temperature $T$, (2) have a high dimensionless sensitivity (logarithmic derivative) $S = -d\log{R} / d\log{T} = -(T/R)dR/dT$ over a wide range of temperature, to resolve small changes in temperature $\Delta T$ over the full range of measurement,  (3) be small and thin, to minimize heat capacity and maximize speed of thermal response, (4) be stable at high temperature, to not lose calibration through general handling and temperature cycling, and (5) have a negligible magnetoresistance, so that  changes in temperature can be distinguished from changes in field. Like all low-temperature thermometers, they also have a negative temperature coefficient (NTC)  $dR/dT < 0$ so as to remain sensitive as low temperature. 
The most commonly used NTC resistive thermometers reported in the literature -- thick-film ruthenium oxide  \cite{koppetzki_thick_1983, li_thick_1986, goodrich_magnetoresistance_1998}, thin-film metal oxynitrides \cite{courts_standardized_2014,  lin_ntc_2018} such as the commercially available Cernox\textsuperscript{\texttrademark} ZrO$_x$N$_y$ \cite{swinehart_metal_1994}, and thin amorphous/polycrystalline films of  Au\textsubscript{x}Ge\textsubscript{1-x} \cite{dodson_metal-insulator_1981, zhu_fabrication_1993, bethoux_au-ge_1995, fortune_physical_1998, dann_au_2019} -- all exhibit some of these qualities, but  none satisfy all of these requirements over the full range of  temperatures ($<100$\,mK  to 300\,K) and dc magnetic fields (0 - 45 T) available 
at sites like the National High Magnetic Field Laboratory (NHMFL) \cite{boebinger_gregory_s_national_2021}. 

Thick film ruthenium oxide sensors prepared by
screen printing of conductive paste containing ruthenium oxide and
bismuth ruthenate have a smaller magnetoresistance  than most other alternatives below 4 K \cite{koppetzki_thick_1983, li_thick_1986, goodrich_magnetoresistance_1998, fortune_high_2000} and  dimensionless sensitivities between 0.25 and 0.5 at 1 K \cite{li_thick_1986}   
but  are largely insensitive at temperatures above 20 K \cite{koppetzki_thick_1983, ihas_low_1998}. They also have a larger mass,  slower thermal response, higher heat capacity,  and poorer thermal  contact than competing thin film sensors. Thin film versions designed to overcome these disadvantages exhibit much lower dimensionless sensitivity ($\lesssim 0.1$),  making them unsuitable for high resolution calorimetry ~\cite{nelson_thin_2015}.

Thin-film metal oxynitride MO$_x$N$_y$ resistors  (where M = Hf, Nb, Ti, Ta, and/or Zr) 
offer higher dimensionless sensitivities (ranging from $0.5 \sim  2$) useful over a much wider temperature range than RuO$_x$ sensors.  These sensors are grown by reactive sputtering of an elemental metal in a controlled mixture of oxygen and nitrogen gas \cite{swinehart_metal_1994, lin_ntc_2018}. Unfortunately, they are also highly magnetoresistive below 4 K ---  approaching 100\% in a field of 15 T at 0.1 K \cite{fortune_high_2000, rosenbaum_anomalous_2001}.

Annealed  polycrystalline films of Au\textsubscript{x}Ge\textsubscript{1-x}  offer low mass,  rapid thermal response, and a high and tunable dimensionless sensitivity ranging from 0.5 to 2 between 30 mK and 300 K  \cite{bethoux_au-ge_1995, fortune_physical_1998} after annealing at  150 \textasciitilde 160$\,^{\circ}$C, 
 making them an attractive choice for zero-field nanocalorimetry \cite{tagliati_differential_2012}.  They can also be readily fabricated through flash evaporation or co-sputtering. The temperature dependence and sensitivity are, however,  strongly dependent upon annealing conditions, as these affect the size and distribution of Au islands formed at the polycrystalline Ge grain boundaries  \cite{fortune_physical_1998}. Films annealed at these temperatures are therefore subject to shifts in calibration 
 \cite{fortune_physical_1998}. Also, these films exhibit significant magnetoresistance at low temperatures, reaching 20\% in a 20 T field at 0.1 K \cite{fortune_high_2000}. Annealing at much higher temperatures (450 $^{\circ}$C) in a reducing atmosphere has recently \cite{dann_au_2019} been reported to yield a stable, reproducible sensor with a logarithmic temperature dependence, but the reduced sensitivity  ($S < 0.1$) makes it impractical for high resolution calorimetry.

Less commonly used resistive thermometers include the ceramic-metal composites (cermets) \cite{abeles_granular_1976, gershenfeld_percolating_1988, unruh_granular_1990} re-investigated here. Cermets are produced by the co-sputtering of a metal and an insulator, which produces weakly connected metallic islands in an insulating ceramic sea at metal concentrations close to the percolation threshold. Conduction primarily occurs due to thermally assisted hopping of charge carriers from island to island and is therefore strongly temperature dependent at low temperatures \cite{RevModPhys.79.469}. 
The relatively low magnetoresistance reported for several cermets in magnetic fields above a few tesla \cite{gershenfeld_versatile_1987, gershenfeld_percolating_1988, unruh_granular_1990, patterson_transport_1992} has led us to take a second look at them for thermometry and small sample calorimetry. 
Cermets can be prepared from several elemental metals combined with insulators such as SiO$_2$ and Al$_2$O$_3$ \cite[and references therein]{abeles_granular_1976}. 

The temperature dependence of resistance for a cermet is controlled by the distribution of island sizes, inter-island distances, and metallic  percentage \cite{sheng_feature_1992, RevModPhys.79.469}, providing an opportunity for fine-tuning the sensitivity through co-sputtering in a manner not available using standard deposition methods for thick-film RuO$_x$ and thin film oxynitride MO$_x$N$_y$ resistors. Observations of a pronounced low-field negative magnetoresistance \cite{gershenfeld_versatile_1987} and reduced sensitivity \cite{gershenfeld_percolating_1988, unruh_granular_1990} at low temperature have, however, limited their use for calorimetry and related measurements in magnetic fields \cite{fortune_specific-heat_1990}. These drawbacks turn out to arise from the  particular materials used. Here we switch from elemental metals to low-magnetoresistance metallic alloys and show that co-sputtering of such alloys 
with SiO$_2$ in a reactive oxygen + inert gas mixture produces resistive thin film thermometers with both a high sensitivity over an extended temperature range and a significantly reduced magnetoresistance at low temperatures.

\section{METHODS}
\subsection{Thin-film fabrication}

In a first set of depositions, thin-film cermets were produced by the co-sputtering of nichrome (Ni$_{0.8}$Cr$_{0.2}$, abbreviated here as NiCr) and SiO$_2$ targets onto an insulating  150\,nm thick silicon nitride coated silicon substrate at room temperature (1000\,nm silicon oxide coated silicon substrates produce equivalent results). Prior to deposition, the chamber was pumped down to a base pressure of $<2 \cdot 10^{-8}$\,Torr.  The films were co-sputtered in a working gas of Ar at 3\,mTorr and a flow rate of 25\,sccm. Typical deposition rates were $0.8$--$1.8$\,nm/min for NiCr and $0.5$--$0.6$\,nm/min for SiO$_2$. Completed thin films had a thickness of $50$--$60$\,nm. The final composition was labeled by the NiCr deposition fraction. 

In a second set of depositions, a  dilute mixture of Ar:O$_2$ was substituted for pure Ar as the working gas.   Typical deposition rates were  $1.8$--$2.5$\,nm/min for NiCr and $0.5$--$0.8$\,nm/min for SiO$_2$ producing thin films on the order of $50$--$60$\,nm.  A protective cap of 15\,nm thick SiO$_2$ was deposited on top of the cermet films to prevent any subsequent reaction with the atmosphere.  
The depositions 
were done at a deposition pressure of 3 mTorr and a flow rate of 25 sccm Ar, with O$_2$ flow rates varying from $0.10$--$0.18$\,sccm.  
Before deposition, the chamber was pretreated with a mixture of Ar:O$_2$ gas (4:1 flow rate ratio, 30\,mTorr, 5 min. at room temperature).  This pretreatment occurred after the chamber reached its base pressure of $<2 \cdot 10^{-8}$\,Torr, and just prior to moving the sample from the load lock into the deposition chamber. 

We note that base pressures and control over pretreatment conditions for different sputtering systems have significant influence on the quality of the deposited films and their resulting temperature dependence.  Results are strongly detrimentally affected by the presence of residual gases and/or oxygen-gettering materials left on the chamber walls from prior depositions. We found that an ultra-high vacuum and careful pretreatment of the sputter chamber (as described above) were needed to obtain reproducible results.

\subsection{TEM}

Plan-view TEM samples were prepared using well-known methods \cite{palmer_evolution_1989, Romano_1989}.  Briefly, 3\,mm diameter disks were cut from cermet films deposited on SiO$_2$-coated 280\,$\mu$m thick silicon substrates.  A hand polishing tool was used to polish the backside of the disks to a thickness of about 100\,$\mu$m, and a dimple grinder was used to polish the center of the disk to a thickness of ${\sim}5$\,$\mu$m. The samples were then placed in an ion mill (3\,keV argon, 0.1\,mA, at ${\sim}10^{\circ}$ glancing incidence) and thinned from the backside until a small hole formed in the center.  Images were made in the electron transparent regions adjacent to the hole using a JEOL JEM400 Transmission Electron Microscope operated at 100 kV.  Particle size was estimated by drawing a line of known length across the photo and dividing the length by the number of particles along the line.  This was done three times on each photograph and the average was reported as the estimated particle size.  

\subsection{Resistance measurements}

All resistance measurements were performed at a temperature-dependent power (about 10 fW at 0.1 K, 1 pW at 1 K, and 10 nW at 100 K) chosen to avoid systematic errors due to self-heating. 
Magnetoresistance measurements require a constant temperature while sweeping magnetic field; this was done by integrating the studied thermometer into a membrane-based nanocalorimeter \cite{tagliati_differential_2012}. Constant temperature was then obtained through a local temperature offset heater at constant total heater power, raising the calorimeter thermometer temperature above that of the cryostat bath. Measurements up to 12\,T were taken in a dilution refrigerator with a stable base temperature below 10\,mK (dilution unit in a compensated magnetic field). 
The weak thermal coupling between the studied thermometer on the calorimeter and the thermal bath provides a very stable temperature that is insensitive to magnetic fields. For temperatures above 50\,mK, any magnetic field effects on the cryostat will not affect the temperature on the calorimeter. A stable temperature can thus be obtained by measuring and stabilizing the total power supplied to the calorimeter area and assuming that the calorimeter thermal link is magnetic field independent. Measurements at low calorimeter power with insignificant offset heating but with stabilized base temperature up to 4\,K confirm this assumption.
Measurements at higher fields were taken at NHMFL in the same way using a $^{3}$He system. Magnetic field sweeps below 1\,K were performed 
at very slow ramp rates (below 0.2\,mT/s).

\section{Results}
\subsection{Temperature dependence}

\begin{figure}[t!]
\includegraphics[width = 0.45\textwidth]{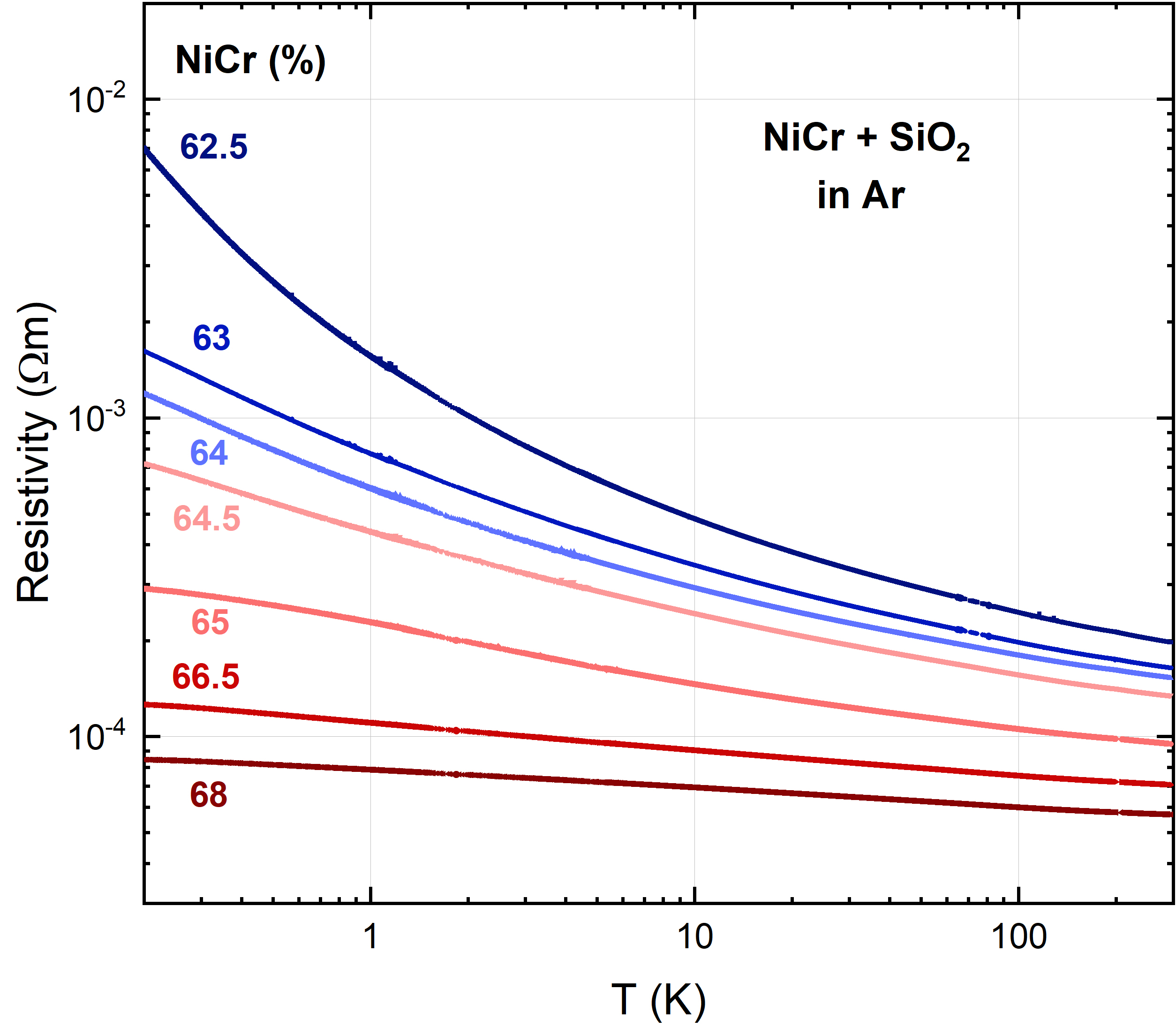}
\caption{Temperature dependent resistivity as a function of NiCr percentage by volume for  (NiCr, SiO$_2$) cermets co-sputtered in argon at 3 mTorr.
}
\label{fig:rho(T)_no_O2}
\end{figure}

The  temperature dependent resisistivities $\rho(T)$ (at zero field) for the first set of depositions (in argon) are shown in Fig.~\ref{fig:rho(T)_no_O2}. 
Importantly, there is a crossover in the slope of the low temperature dimensionless sensitivity from positive to negative  with increasing metal percentage, with a nearly constant dimensionless sensitivity of 0.1 at 66.5\% NiCr metal fraction. This indicates that our samples span the transition region between insulating and metallic behavior \cite{mobius_metal-insulator_2018}. Both the resistivity and sensitivity increase with decreasing temperature for insulating samples, with the 62.5\% sample representing the lower limit for metal concentration for this particular co-sputtering method (due to the large difference in metals and insulator sputtering rates).  

\begin{figure}
\includegraphics[width = 0.45\textwidth]{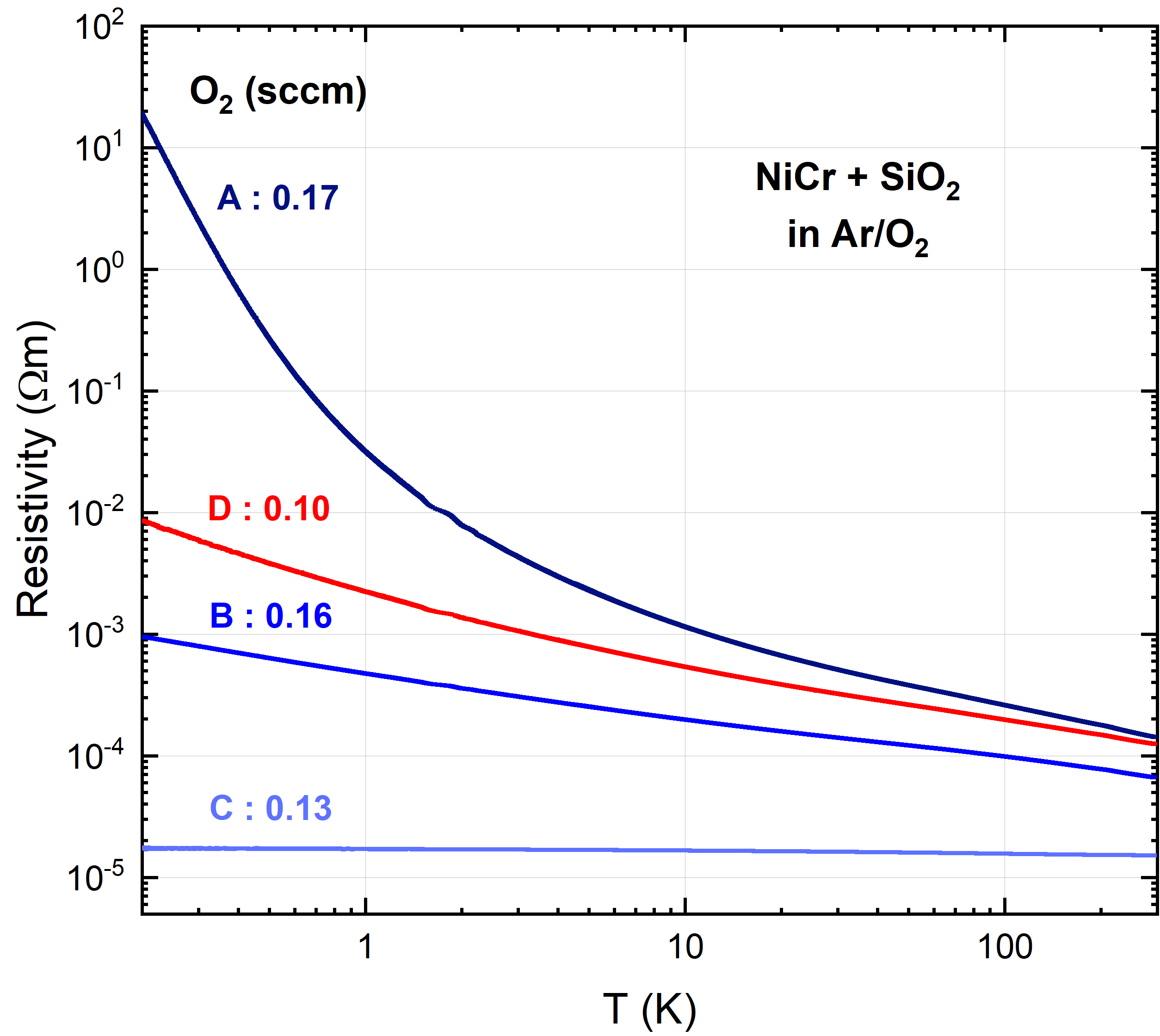}
\caption{Measured resistivities $\rho(T)$ for different oxygen flow levels during co-sputtering. Increasing the oxygen flow rate induces a transition from metallic to insulating behavior. Films A, B, and C were deposited at a NiCr rate of 2.81 nm/min. and SiO$_2$ rate of 0.78 nm/min. with oxygen flow rates of 0.17, 0.16, and 0.13 sccm respectively.  Film D was deposited at a NiCr rate of 1.99 nm/min. and SiO$_2$ rate of 0.58 nm/min. with an oxygen flow rate of 0.10 sccm. In all cases the argon working gas flow rate was 25 sccm.  The dimensionless sensitivity of film D is nearly ideal for a wide-range sensor, increasing from 0.5 at 100 K to 1.0 at 0.2 K.}
\label{fig:rho(T) for O2}
\end{figure}

In Fig.~\ref{fig:rho(T) for O2}, the results from the second set of depositions in an Ar:O$_2$ atmosphere are shown for a set of three films (A, B, and C) deposited  at constant NiCr and SiO$_2$ sputtering rates but with differing oxygen flow rates. Also shown is a film D deposited at a lower metal sputtering rate using an oxygen flow rate of 0.10 sccm. Sample C is metallic, while the other three represent the crossover to insulating behavior. 
The large changes in temperature dependence of the resistivity and sensitivity between samples A, B,  and C produced by small changes in oxygen flow rate from 0.13 to 0.17 sccm illustrate the sensitive dependence of thermometer properties on oxygen concentration.  The relative effect of changes in the metal sputtering rate and oxygen flow rate is illustrated by sample D, deposited at both lower oxygen flow and lower metal deposition rate.  An increase in oxygen flow rate thus contributes to the crossover from metallic to insulating behavior in a manner similar to an increase in the SiO$_2$ sputtering rate. 

These as-deposited films --- capped by a thin insulating layer of SiO$_2$ to protect against humidity and further oxidation --- are remarkably stable. To within the resolution of our measurements, no shifts in calibration were seen in  samples remeasured 5 years later. With the exception of the highest resistivity samples (such as sample A in Fig.~\ref{fig:rho(T) for O2}), no shifts in calibration were observed after annealing for one hour at 350 $^{\circ}$C nor after an additional 10 min at 380 $^{\circ}$C. For the highest resistivity samples, a small decrease in resistivity is sometimes seen at low temperatures after a first anneal, but no further changes are seen with additional anneals.

\subsection{TEM}
\begin{figure*}
\includegraphics[scale = 0.205]{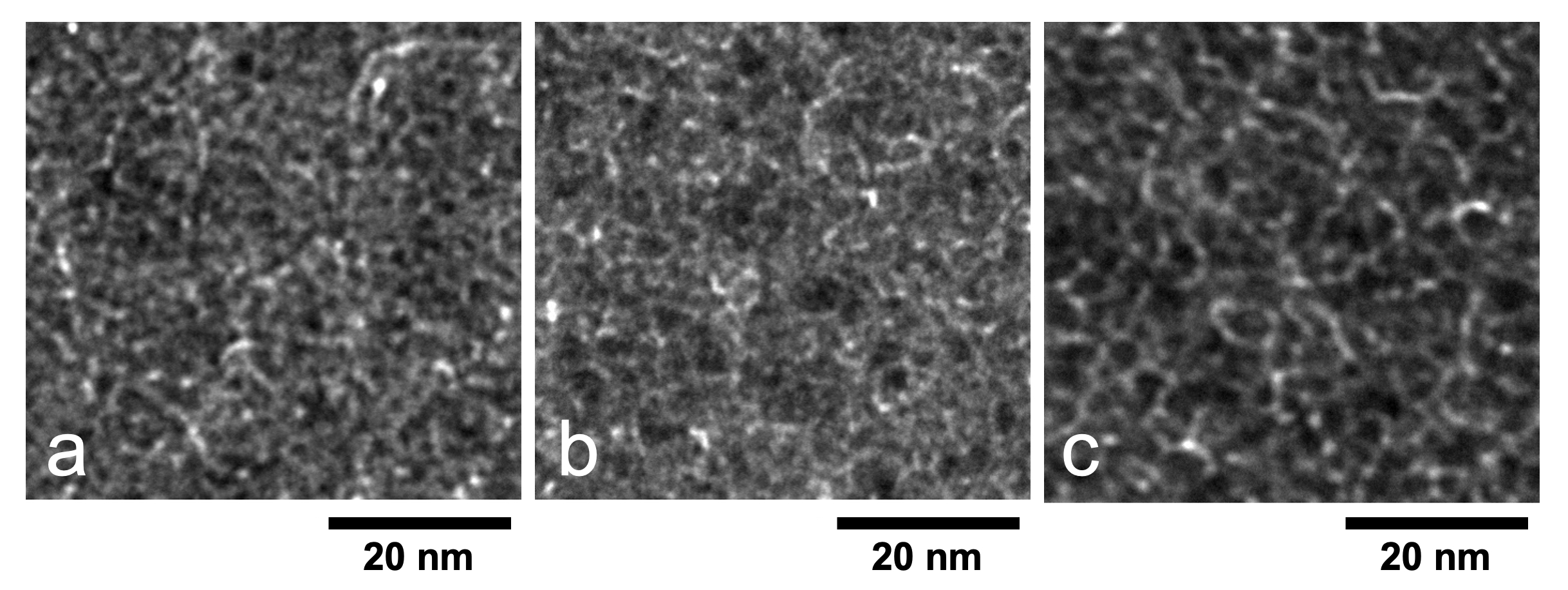}
\caption{Plan-view TEM micrographs of cermet films grown with varying O$_2$ flow conditions producing (a) higher than optimal, (b) optimal, and (c) lower than optimal $\rho$ vs T characteristics.  Larger metal particle sizes correlate with lower sensitivity. All films are co-sputtered in Ar:O$_2$ at a working pressure of 3\,mTorr with an argon flow of 25\,sccm. 
}
\label{fig:islandsize_as_O2_rate_changes}
\end{figure*}

\begin{figure*}
\includegraphics[scale = .18]{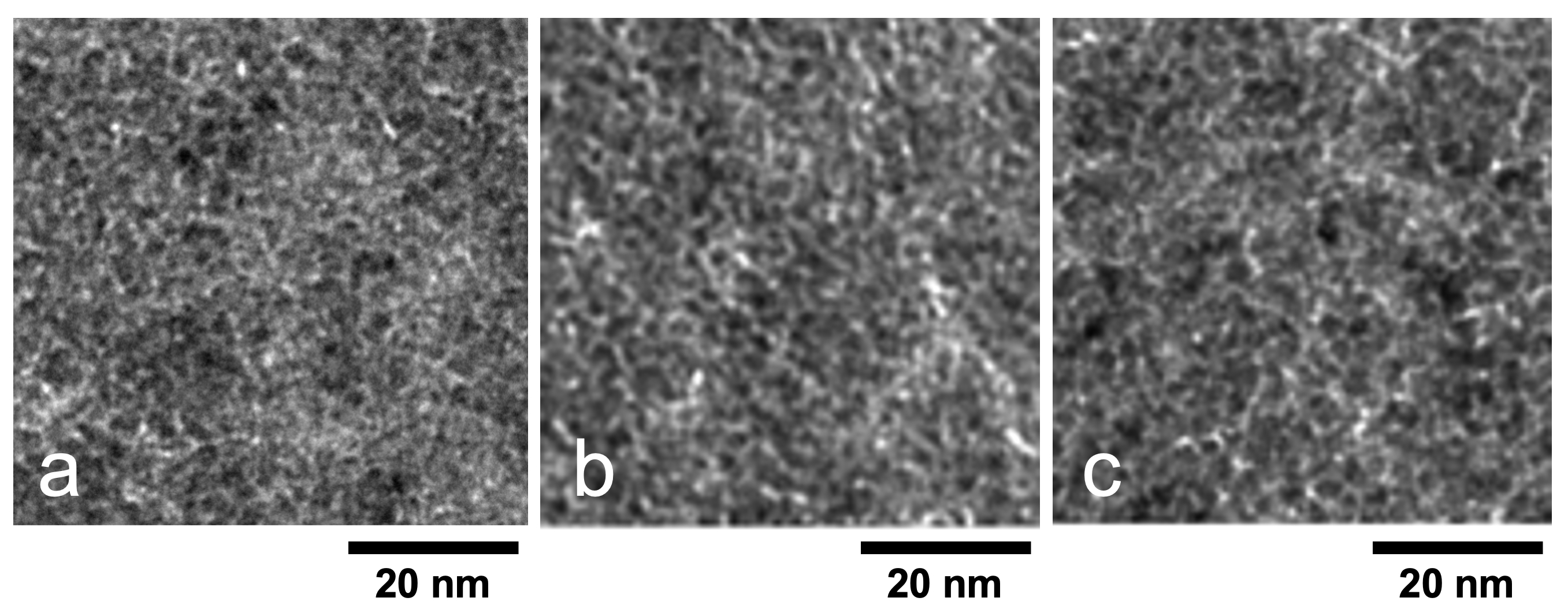}
\caption{Plan-view TEM micrographs of cermet films grown with varying metal sputtering rates producing (a) higher than optimal, (b) optimal, and (c) lower than optimal $\rho$ vs T characteristics.  Films with higher metal rate (keeping the same O$_2$ flow rate and SiO$_2$ sputtering rate) leads to an increase in average metal island size. All films are co-sputtered in Ar:O$_2$ at a working pressure of 3\,mTorr with an argon flow of 25\,sccm.
}
\label{fig:islandsize_as_NiCr_rate_changes}
\end{figure*}

Plan-view TEM micrographs in Fig.~\ref{fig:islandsize_as_O2_rate_changes} show cermet films grown at constant NiCr and SiO$_2$ sputtering rates but with differing oxygen flow rates. The micrographs indicate that the change to more insulating properties arises from a decrease in average metal island size with increasing oxygen concentration.
Figure~\ref{fig:islandsize_as_NiCr_rate_changes} shows TEM micrographs as a function of increasing metal/insulator sputtering rate ratio at constant oxygen concentration. The metal island size at any given oxygen flow rate is seen to increase with increasing metal-to-insulator sputtering rate ratio.
Samples grown without O$_2$ show the same trend --- increasing island size with greater metal content --- but the island sizes tend to be larger when sputtered without oxygen, with sizes ranging from 4.0\,nm for the 62.5\% sample in Fig.~\ref{fig:rho(T)_no_O2} to 5.0\,nm for the 68\% sample.

\begin{table}[hbt! ]

\caption{\label{table:tem fig params} Film Growth Parameters for Fig.~\ref{fig:islandsize_as_O2_rate_changes}  and Fig.~\ref{fig:islandsize_as_NiCr_rate_changes}.}

\begin{ruledtabular}

\begin{tabular}{cccccc}

Figure & NiCr rate & SiO$_2$ rate & $R_\mathrm{RT}$ & O$_2$ flow & Diameter\\
 & (nm/min) & (nm/min) & $({\mathrm{k}\Omega})$ & (sccm) & (nm) \\
\tableline
~\ref{fig:islandsize_as_O2_rate_changes}a & 2.10  & 0.58 & 30-35 & 0.15 & 3.1\\
~\ref{fig:islandsize_as_O2_rate_changes}b & 2.10  & 0.58 & 1.2-1.3 & 0.12 & 4.4\\
~\ref{fig:islandsize_as_O2_rate_changes}c & 2.10  & 0.58 & 0.43-0.48 & 0.10 & 5.7 \\
~\ref{fig:islandsize_as_NiCr_rate_changes}a & 2.60  & 0.80 & 2.0-2.8 & 0.19 & 2.8 \\
~\ref{fig:islandsize_as_NiCr_rate_changes}b & 2.74  & 0.80 & 1.5-2.0 & 0.19 & 3.2 \\
~\ref{fig:islandsize_as_NiCr_rate_changes}c & 2.95  & 0.80 & 0.75-0.85 & 0.19 & 3.6 \\
\end{tabular}
\end{ruledtabular}
\end{table}

\subsection{Magnetic field dependence}

\begin{figure}
  \centering
\includegraphics[width=1.025\linewidth]{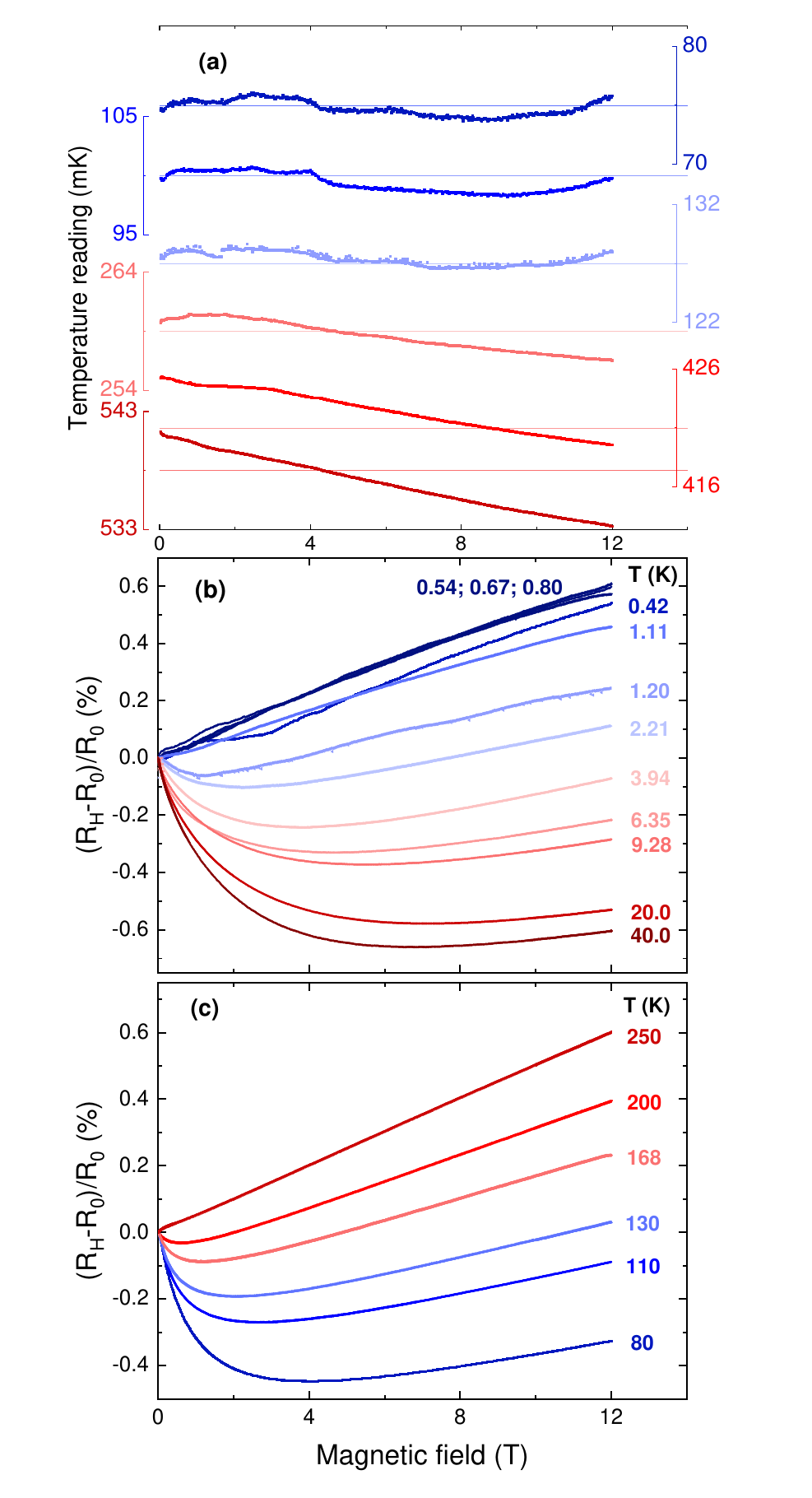}
  \caption{Magnetoresistance as a function of magnetic field up to 12 T, at various temperatures for (NiCr, SiO$_2$) films co-sputtered in O$_2$ and Ar at 3 mTorr pressure. The thermometer films have room temperature resistivities around 7$\times 10^{-5}\,\Omega \mathrm{m}$, and show a temperature dependence similar to film B in Fig.~\ref{fig:rho(T) for O2}. (a) Measured temperature within 10\,mK windows. Below 150 mK, the effect of magnetic field sweep on actual temperature affects the accuracy of magnetoresistance.
The magnetoresistance at higher temperatures, (b) and (c), is composed of a negative, low-field magnetoresistance and a linear, positive high-field magnetoresiance.}
\label{fig:MagRes}
\end{figure}

\begin{figure}
  \centering
\includegraphics[width=1\linewidth]{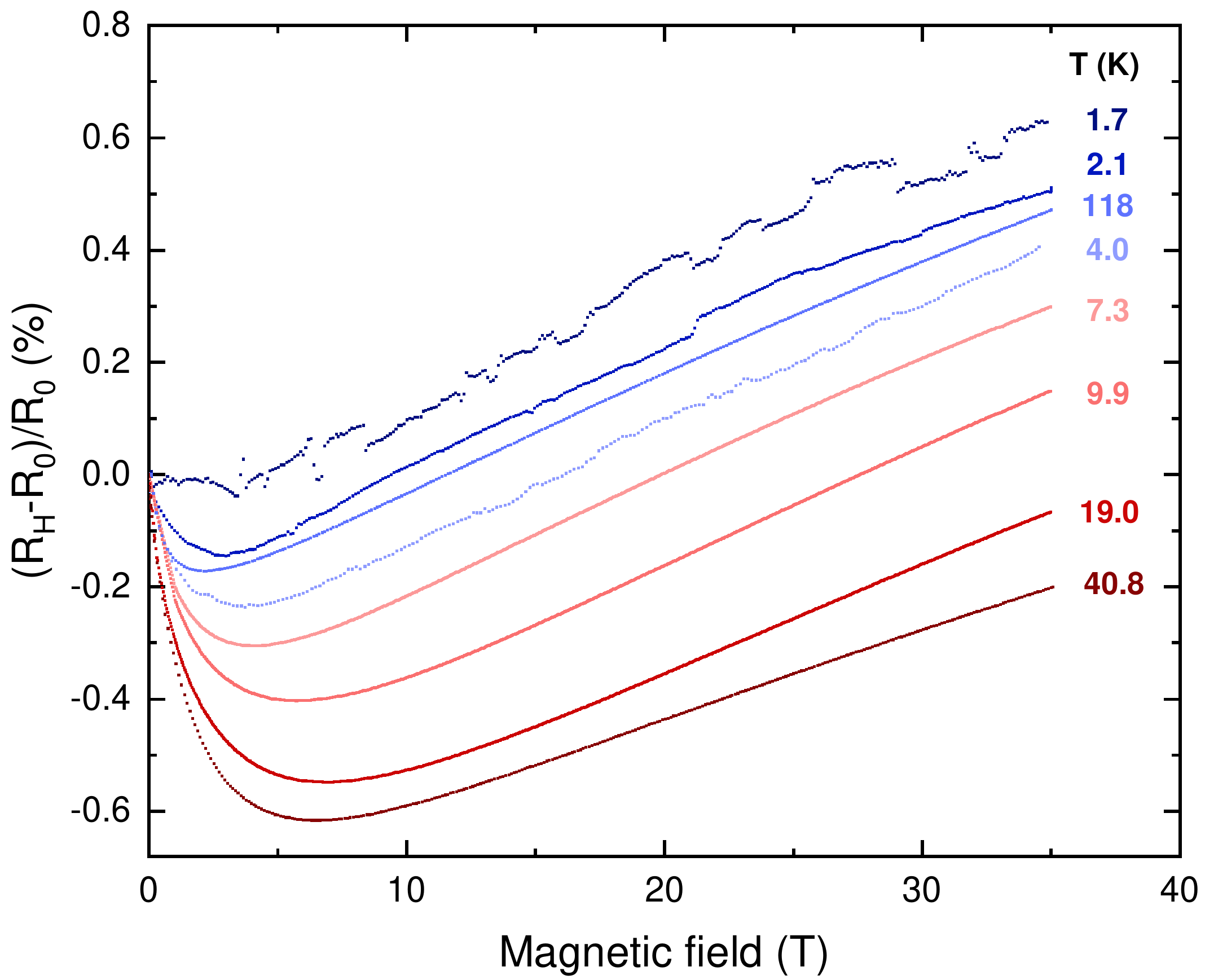}
  \caption{Magnetoresistance as a function of magnetic field up to 35\,T at various temperatures, expressed as relative resistance change. The curve at 1.7\,K is affected by electronic noise sporadically changing temperature by a few mK. The thermometer film has a room temperature resistivity value of 9.17$\times 10^{-5}\,\Omega \mathrm{m}$, and the temperature dependence curve lies between films B and D in Fig.~\ref{fig:rho(T) for O2}.
  }
\label{fig:HighField}
\end{figure}

In Fig.\,\ref{fig:MagRes}, the thermometer magnetoresistance is shown for magnetic fields up to 12\,T and temperatures from 75\,mK to 250\,K. At low fields, the magnetoresistance is negative except for the lowest temperatures, $T<1.1$\,K, and  highest temperatures $T\gtrsim 230$\,K. The size of the negative magnetoresistance term increases with decreasing temperature down to 40\,K, see Fig\,\ref{fig:MagRes}(c), then becomes smaller again at still lower temperatures, see Fig.\,\ref{fig:MagRes}(a,b). Near room temperature, the magnetoresistance is positive and linear with field.
The magnetoresistance at high magnetic fields increases linearly with field at all temperatures as seen in Fig.~\ref{fig:HighField}. 
For the studied samples, the zero-field dimensionless sensitivity varies between 0.28 and 0.38 over the temperature range. The maximum effect of magnetic field on apparent temperature at any temperature is only 2\%.

\section{DISCUSSION}

\subsection{Temperature dependence}

One useful physical model \cite{mobius_metal-insulator_1999, mobius_metal-insulator_2018} for the temperature dependent resistivity $\rho(T)$ of materials near the metal-insulator transition involves a crossover from variable-range hopping at the lowest temperatures to a nearly constant ``disordered film'' resistance at high temperature.  We represent this here  by a phenomenological  extended variable-range hopping model incorporating a crossover from stretched exponential to augmented power law dependence \cite{mobius_metal-insulator_2018} for films close to the metal-insulator transition: 
\begin{equation}
   \label{eq:Mobius}
    \rho(T) = \rho_0 \left[ 1 +\left(\frac{T_0}{T}\right)^{\nu} \right] e^{({T_0}/{T})^{\nu}} 
\end{equation}
where $k_B T_0$ represents a characteristic energy for variable-range hopping between adjacent metal islands in an insulating medium,  ${\nu}$ is a measure of power law sensitivity whose value depends (in part) on the distribution of effective distances, and $\rho_0$ is a scaling parameter equal to the  residual resistivity in the $T \rightarrow \infty$ high temperature limit. 
$\rho_{0}$, $T_0$,  and $\nu$ all depend on metal island percentage $x$.

We find that $\rho(T)$ curves in Fig.~\ref{fig:rho(T)_no_O2}  for (NiCr, SiO$_2$) cermets co-sputtered in a pure argon atmosphere are well described by Eq.~(\ref{eq:Mobius}) for NiCr percentages near a metal-insulator transition at $x_{\mathrm{MIT}}$; the dependence of $\rho_0$ on metal concentration $x$ can be modeled in terms of percolation theory \cite{McAlister1985}. Details and limitations of the modeling are  discussed in the appendix. 

Experimentally, the residual resistivity $\rho_{0}$, power-law coefficient $\nu$, and hopping temperature $T_0$ all decrease  with increasing metal percentage. As a result,  materials described by this model exhibit a tradeoff:  increasing sensitivity $\nu$ leads to higher residual resistivity $\rho_{0}$.

There is a notable resemblance between the temperature dependence shown in Fig.~\ref{fig:rho(T)_no_O2} and that seen for 100\,nm thick Ni$_x$(SiO$_2$)$_{1 - x}$ cermets \cite{unruh_granular_1990} co-sputtered in argon for atomic percentages ranging from $x = 0.6$ to $x = 0.9$, having Ni metal island sizes ranging from 5 to 10\,nm for insulating films. 

We observe a room temperature resistivity of $200\, \mu \Omega$m at 62.5\%\,NiCr, while the reported value for 66\% Ni cermet with a comparable temperature dependence is $670\, \mu \Omega$m \cite{unruh_granular_1990}. Comparable sensitivities lead to similar room temperature resistivities, albeit at measurably higher metal percentages for the Ni-based than NiCr based materials.  
A main difference, however, is in the low-temperature sensitivity, which abruptly decreases below 2\,K  for the Ni-based cermets. This decrease in sensitivity around 2\,K was also seen for Pt-based cermets \cite{gershenfeld_percolating_1988}. A second difference is the larger effect of magnetoresistance at low temperature when using the Ni-based films as thermometers \cite{unruh_granular_1990}.  

Our results shown here in Fig.~\ref{fig:rho(T)_no_O2} --- in particular the crossover from metallic to strongly temperature dependent insulating behavior and low temperature magnetoresistance--- also bear some resemblance to results observed between 1 K and 4 K for thin metal film thermometers formed from a film consisting of discontinuous metal islands of NiCr sandwiched between two thick SiO$_x$ layers \cite{griffin_low_1974}. The temperature dependence above 4 K was not reported, however, as the films studied in Ref.~\cite{griffin_low_1974}, are unstable at room temperature, typically increasing in resistance by an order of magnitude after 30 days.  

\subsection{Dimensionless sensitivity}

\begin{figure}
  \centering
\includegraphics[width=1\linewidth]{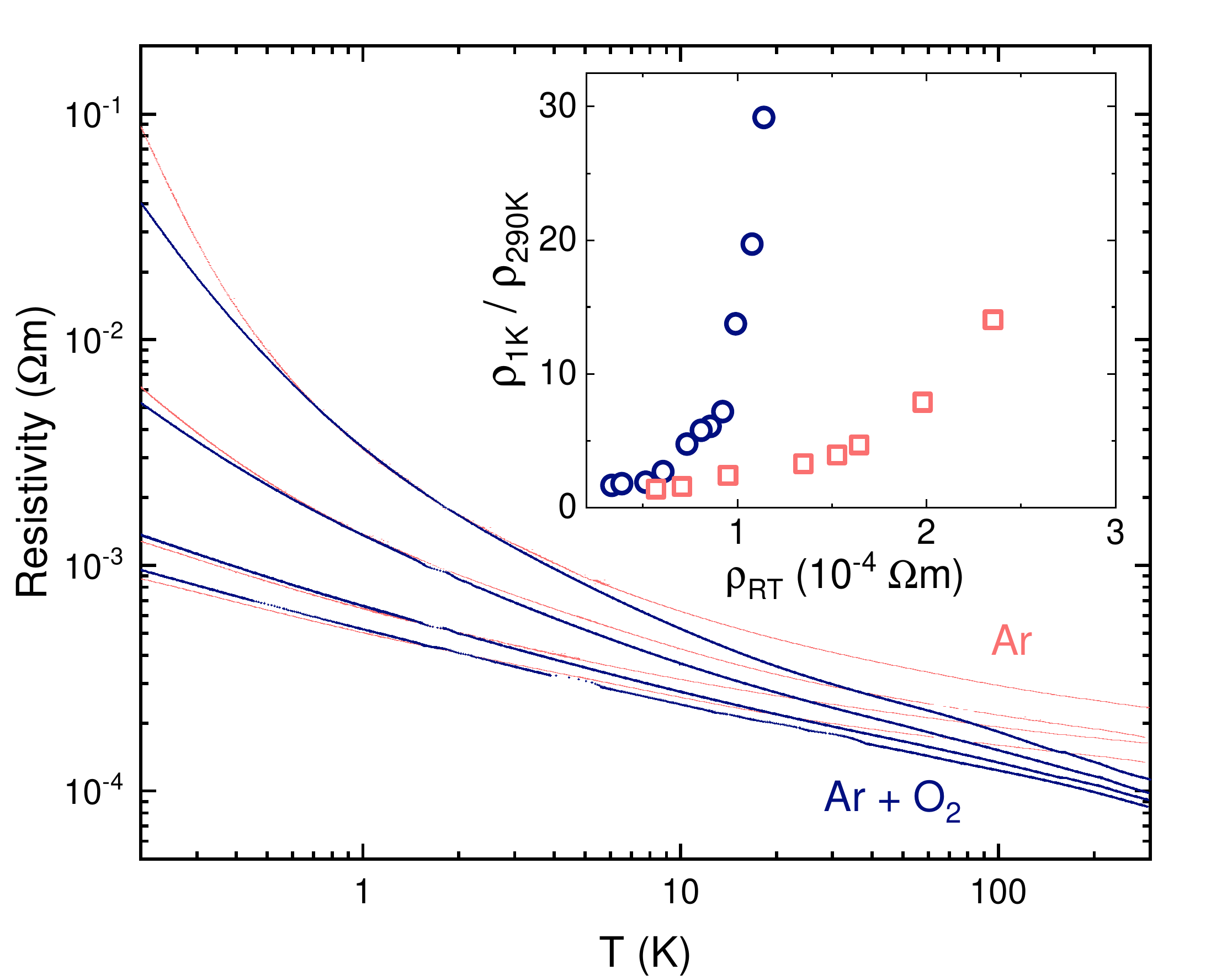}
  \caption{Comparison of temperature-dependent resistivities of (NiCr, SiO$_2$) films fabricated in both Ar and Ar + O$_2$ environments. Curves were selected to match at about 1\,K. Co-sputtering in a reactive mixed Ar + O$_2$ atmosphere yields films with significantly greater sensitivity between 10 K and room temperature. The inset describes the resistivity ratio as a function of room temperature resistivity for the two fabrication methods. Materials in the same fabrication family obey a common curve, allowing a quick comparison of materials. }
\label{fig:RRCompare}
\end{figure}

Unlike the films sputtered in a pure Ar atmosphere (Fig.~\ref{fig:rho(T)_no_O2}), the oxygen grown films (Fig.~\ref{fig:rho(T) for O2}) retain a higher sensitivity even at room temperature, combined with an even lower room temperature resistivity, making these films suitable for calorimetry at temperatures up to and including room temperature. For film D in Fig.~\ref{fig:rho(T) for O2}, the dimensionless sensitivity increases from a minimum of 0.5 at 100\,K to a maximum of 1.0 at 0.2\,K, a sensitivity comparable to that seen in commercially available thin-film metal oxynitride ZrO$_x$N$_y$ resistors \cite{courts_standardized_2014}.
In Fig.~\ref{fig:RRCompare}, curves with comparable resistivity at 1~K, fabricated with Ar only and Ar + O$_2$ atmosphere, are compared. As seen in the figure, the room temperature resistivity is significantly lower for the Ar + O$_2$ grown samples.
Within each preparation method, we also see that we can infer the low-temperature resistance of any particular (NiCr, SiO$_2$) film from its room temperature value. This implies that two identically prepared films with the same room temperature resistivity should follow a common curve, as illustrated in the inset of Fig.~\ref{fig:RRCompare}.

Sputtering in a reactive oxygen atmosphere can be expected to change the resistive properties of the metal-alloy + insulator cermets in at least two  ways. First, sputtering in a reactive oxygen atmosphere is expected to improve the stoichiometry of the sputtered SiO$_2$ \cite{jones_effect_1968, jeong_characterization_2004} and may prevent forming possible metal silicides. Second, as the oxygen flow rate is increased, some Cr in Ni$_{0.8}$Cr$_{0.2}$ may react to form Cr$_2$O$_3$ due to the preferential oxidation of Cr. Put together, more insulator should be deposited when O$_2$ is present in the working gas. Consistent with this expectation, we find that we can use higher metal deposition rates when using dilute Ar:O$_2$ working gas instead of 100\% Ar to obtain films with comparable room temperature resistivity.

While the effect on cluster size with changing metal/insulator ratio, shown in Fig.~\ref{fig:islandsize_as_O2_rate_changes} and Fig.~\ref{fig:islandsize_as_NiCr_rate_changes} is clear, the net effect for films with similar low temperature behavior is rather subtle. Films grown in Ar behave similarly to those grown in Ar + O$_2$. We find slightly larger island sizes for optimal films in the absence of oxygen, but the effect is weak, less than $\sim 1$\,nm. A more likely explanation for the effect of oxygen is that the barriers between clusters are improved, affecting the exponent $\nu$ of Eq.~(\ref{eq:Mobius}), and possibly increasing the metal/insulator ratio at the point of optimum low-temperature behavior, thereby lowering the high-temperature resistivity. 

\subsection{Magnetoresistance} 

For thermometry and calorimetry, lower magnetoresistance is always helpful, but it is the magnitude of the magnetic-field induced effect on temperature reading rather than the magnetoresistance itself that is the most important consideration.
For the (NiCr, SiO$_2$) sensors shown in Figs.~\ref{fig:MagRes} and \ref{fig:HighField}, the magnetic-field-induced fractional change in apparent temperature $\Delta T/ T_0$ is less than 2\% over the entire temperature and field ranges measured (75\,mK -- 300\,K and 0 -- 41\,T).
The low values of $\Delta T/T_0$ observed here are a significant improvement over that typically observed for ruthenium oxide sensors \cite{ihas_low_1998, goodrich_magnetoresistance_1998} ---Ref.~\cite{goodrich_magnetoresistance_1998} reports a 28\% change in $\Delta T/T_0$ at 0.62 K in a 32 T field ---  and cermets of comparable low temperature sensitivity formed from elemental metals \cite{abeles_granular_1976}. 

The positive magnetoresistance component at high fields has a linear field dependence and is nearly temperature independent, thus easily corrected for using methods some of us have presented elsewhere \cite{fortune_high_2000}. The decrease in negative magnetoresistance with decreasing temperature is welcome but unusual. Over the same temperature range, the  magnetic-field induced change in resistance  systematically increases as the temperature decreases for Cernox (TM) and related MO$_x$N$_y$ thin film, AuGe thin film, and RuO$_x$ thick film resistors. An increase is also seen for the (Ni, SiO$_2$) cermet \cite{unruh_granular_1990}.
Understanding of magnetoresistance in granular metals such as the cermets presented here is incomplete and continues to evolve \cite{sheng_feature_1992, aronzon_chapter_2007, RevModPhys.79.469}. That said, the most common explanation for negative magnetoresistance in elemental metals with long mean free paths is the destruction of coherent backscattering due to the application of the magnetic field. Coherent backscattering increases the resistivity of the material at zero field; destroying that coherence by magnetic field leads to a decrease in resistivity.

A main difference between pure metal cermets and cermets based on alloys is the decreased mean free path of the alloy. A short enough mean free path minimizes any coherent backscattering and similar magnetoresistance effects. High resistivity alloys, including NiCr \cite{https://doi.org/10.1002/pssa.2210170217,Samolyuk2018}, are in the Mott-Ioffe-Regel limit \cite{IoffeRegel1960,Mott1972-MOTCIN-10}, characterized by electronic mean free paths comparable to the inter-atomic distance.
Comparing metals and various high resistivity materials such as constantan, nichrome, and phosphor bronze, it is seen that the higher the ratio between room temperature and low temperature resistivity, the larger the magnetoresistance at low temperature \cite{abrecht_magnetization_2007}. Thus, disorder-induced electron scattering by defects and introduced impurities are known to suppress magnetoresistance. Here, the isolated metal grains in combination with using a low-magnetoresistance metal alloy provides a means of producing a useful temperature dependence without introducing high magnetoresistance in a material with a  mean free path that evidently remains in the Mott-Ioffe-Regel limit down to the lowest temperatures.

The role of combining different magnetic $d$-elements for strong disorder scattering at the Fermi level for high entropy alloys such as NiCr is described in detail in Ref.~\cite{mu2019uncovering}. In these materials, scattering can be understood in terms of disorder smearing of the Fermi surface originating from site-to-site potential fluctuations \cite{mu2019uncovering}. Maximum disorder is obtained when there is a large band-center mismatch, such as that obtained by combining nearly filled $d$-band elements (ferromagnetic Fe, Co, Ni) with elements having half-filled $d$-bands (antiferromagnetic Cr, Mn). This suggests that metal combinations that do not scatter strongly in both spin channels at the Fermi level will have a smaller but more temperature-dependent  residual resistivity and, by implication a larger magnetoresistance at the lowest temperatures.

The preferred choice of metal alloy for a particular cermet thermometry application will depend on additional considerations such as the magnetic susceptibility, the magnetic field dependence of the specific heat, and the suitability of the material for use in a sputtering system. In this regard, we have found promising indications that similar results as presented here can be obtained by replacing NiCr with an arc-melted TiCr alloy (7 percent Ti by volume)  sometimes used in calorimetric applications \cite{schwall_automated_1975} as a less magnetic alternative to NiCr for heater elements.    

\section*{conclusions}

In conclusion, we have found that metal-ceramic composite thin films, in particular (Ni$_{0.8}$Cr$_{0.2}$, SiO$_2$), fulfill all requirements expected for thin-film thermometer materials for measurement in high magnetic fields at low temperature. The use of a high resistivity metal alloy  greatly reduces the magnitude of the low field  magnetoresistance compared to that seen for oxygenated cermets formed from elemental metals \cite{gershenfeld_versatile_1987, gershenfeld_percolating_1988}.
The optimized thermometers display monotonic temperature dependence of the resistivity with a dimensionless sensitivity between 0.3 and 0.5, from 300\,K to below 100\,mK. They can be fabricated as stable, square thin film resistors with a predictable low temperature sensitivity, and their magnetoresistance is found to be less than 1\% at all studied temperatures. 
Co-sputtering the films in an atmosphere with a small amount of oxygen extends the useful temperature range of the resistive thin films by simultaneously increasing the sensitivity and decreasing the room temperature resistivity  (compared to sputtering in 100\% Ar). The  room temperature resistivity and sensitivity can be  further fine-tuned by small changes in the metal-to-insulator sputtering rate ratio.

\section*{appendix: extended variable range hopping model}

In this appendix, we test the ability of the extended variable range hopping model \cite{mobius_metal-insulator_2018} introduced to describe the temperature dependence of the resistivity (in the absence of an applied magnetic field). We also present corresponding graphs for the temperature dependent logarithmic sensitivity $S$.  We find that the model describes the temperature dependence of the co-sputtered (NiCr, SiO$_2$) films deposited in a pure argon atmosphere at all temperatures --- even at temperatures well above the characteristic hopping energy temperature $T_0$ --- but fails to fully account for the temperature dependence of films co-sputtered in a mixed argon + oxygen atmosphere. 

\subsection*{Variable-range hopping model}

We start with the expression given by Eq.~(\ref{eq:Mobius}) for the extended variable-range hopping resistivity \cite{mobius_metal-insulator_2018}. 
Expanding this expression in the high temperature limit, using $\delta = (T_0/T)^{\nu} \ll 1$, gives 
\begin{equation}
   \label{eq:Mobius_high_T_limit}
    \rho(T)\bigg\vert_{T\rightarrow \infty}   = \rho_0\left(1+\delta\right)\left( 1 + \delta + \frac{1}{2}\delta^2...\right)\bigg\vert_{\delta\rightarrow 0}  = \rho_0 
\end{equation}
indicating that the scaling parameter $\rho_0$ corresponds to the theoretically expected (constant) residual resistivity. The logarithmic sensitivity $S$ is given by
\begin{equation}
    \label{eq:sensitivity}
    S = - \frac{d\log{R}}{d\log{T}} = - \frac{T}{R}\frac{dR}{dT}= - \frac{T}{\rho(T)}\frac{d\rho(T)}{dT}
\end{equation}
where the resistance $R$ is related to the resistivity $\rho$ as  $R = (L/A)\rho$ 
for a thin film with length $L$, width $w$, thickness $d$, and cross-sectional area $A = w d$.

In calorimetry, the logarthmic sensitivity is important because measurement resolution depends directly on the smallest resolvable temperature change $\Delta T\vert_\mathrm{min}$. This is inversely proportional to $S$ for a given minimum resolvable fractional change in resistance $(\Delta R/R)_\mathrm{min}$:  
\begin{equation}
    \label{eq: T_resolution}
    \Delta T\big\vert_\mathrm{min} = - \frac{1}{S}\left(\frac{\Delta R}{R}\right)_\mathrm{min} T
\end{equation}
Applying Eq.~(\ref{eq:Mobius_high_T_limit}) to Eq.~(\ref{eq:sensitivity}) in the high temperature limit, we find that  
\begin{equation}
    \label{eq:sensitivity_highT}
    S\big \vert_{T > T_0} =  2\nu \left(\frac{T_0}{T}\right)^{\nu}
\end{equation}

In this $T > T_0$ approximation, a constant power-law sensitivity $\nu$ corresponds to an increasing numerical value for the logarithmic sensitivity as $T$ decreases , with $S$ expected to approach a reference  value of $2\nu$ as T approaches $T_0$.

\subsection*{Cermets sputtered in an inert Ar atmosphere}

\begin{figure}[tb!]
\includegraphics[width = 0.45\textwidth]{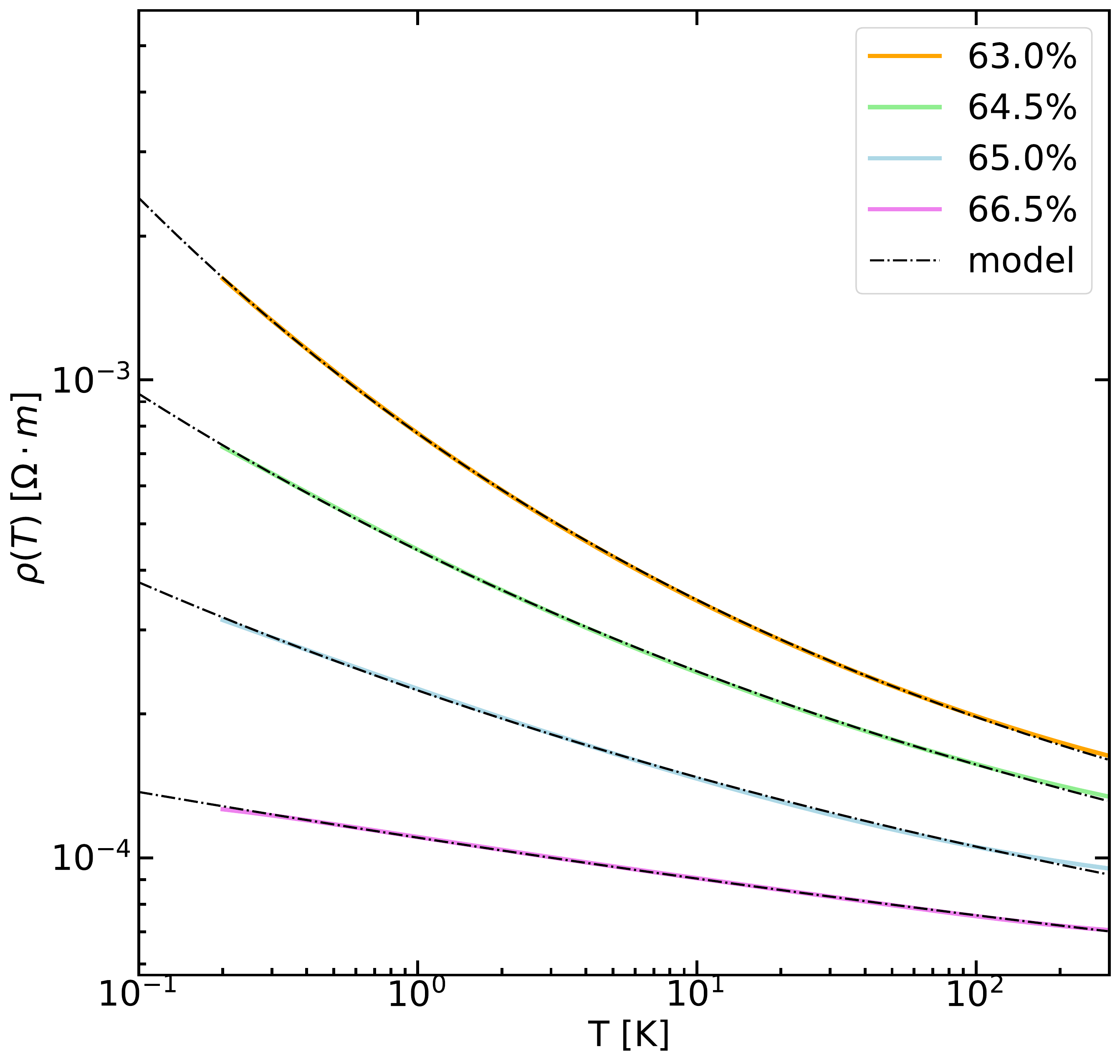}
\caption{Fit of extended variable range hopping model to data for various NiCr concentrations. Solid lines correspond to data and dashed lines to the corresponding model fits to Eq.~(\ref{eq:Mobius}).  
}
\label{fig:Mobiusfit}
\end{figure}

\begin{figure}[t!]
\includegraphics[width = 0.45\textwidth]{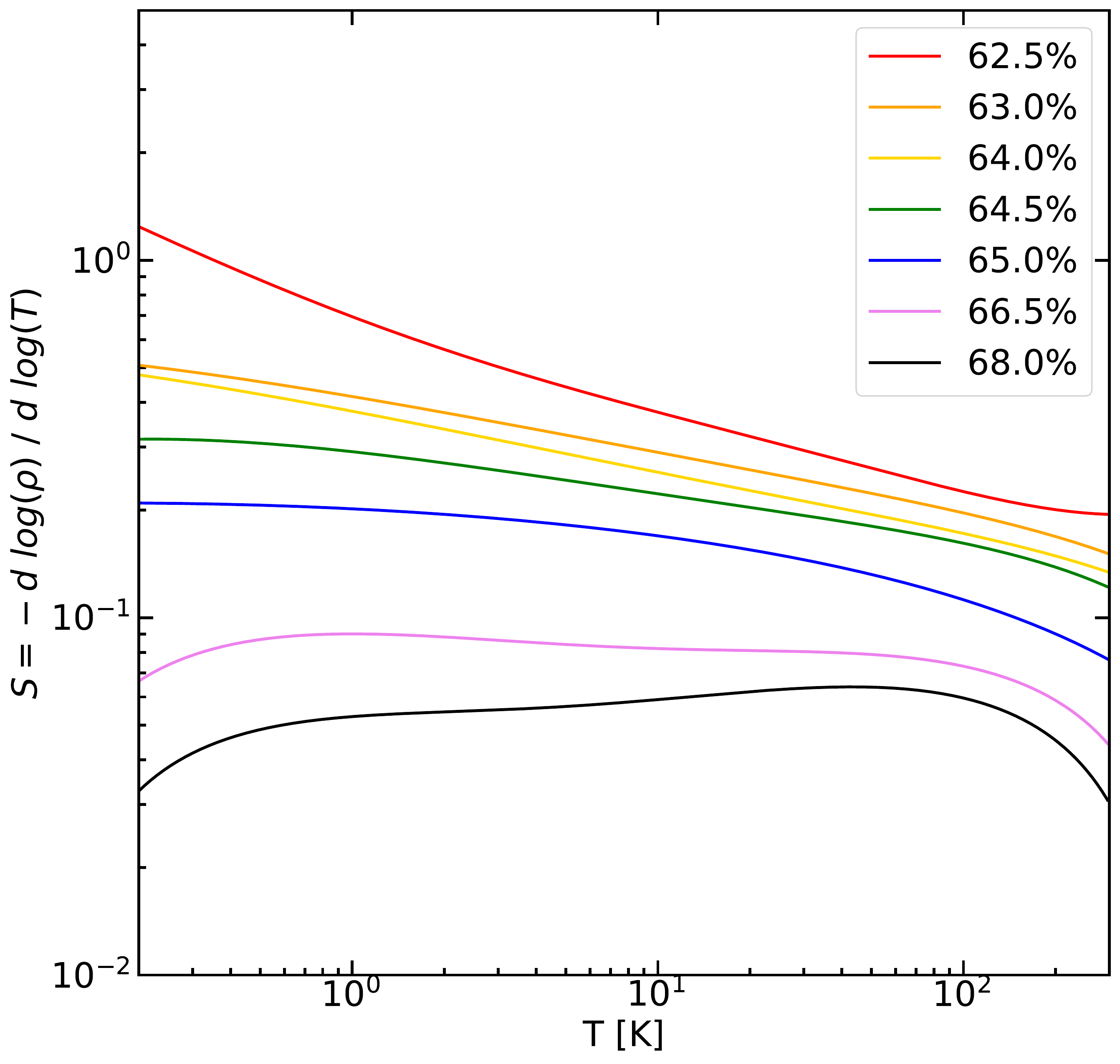}
\caption{Logarithmic (dimensionless) sensitivity $S$ for the films shown in Fig.~(1) of the main text.} 
\label{fig:S_for_Mobiusfit}
\end{figure}

Figure~\Ref{fig:Mobiusfit} shows fits of the empirical variable range hopping model, Eq.~(\ref{eq:Mobius}), to data for various NiCr concentrations. These represent a subset of the data presented in Fig.~1 of the main text for NiCr percentages ranging from 63\% to 66.5\%. As seen in Fig.~\ref{fig:Mobiusfit},  the temperature dependent resistivity $\rho(T)$ of these films can be fit to the model of Eq.~(\ref{eq:Mobius}) over the entire temperature range from 0.2\,K to 300\,K.
The corresponding sensitivities for the data presented in Fig.~1 of the main text are shown in Fig.~\ref{fig:S_for_Mobiusfit}. Note how the slope $dS/dT$ of the sensitivity changes from negative to positive as we cross the metal-insulator transition.
The resulting fit parameters for films on the insulating side of the metal-insulator transition are  listed in Table~\ref{table:VRH model}.

\begin{table}[hbt! ]
\caption{\label{table:VRH model} Dependence of the variable-range hopping fitting parameters found from non-linear least square fit to Eq.~(\ref{eq:Mobius}) on  NiCr metal percentage for (NiCr, SiO$_2$) cermets co-sputtered in pure Argon. }
\begin{ruledtabular}
\begin{tabular}{llll}
metal \% & $\rho_0\,(\mu\Omega\mathrm{m})$      & $T_0\,(\mathrm{K})$ & $\nu$    \\
\tableline
63.0     & 57.0(2)  & 14.9(2) & 0.183(2)  \\
64.5     & 38.1(3) & 22.3(4) & 0.135(4) \\
65.0     & 30.6(4)  & 4.9(4) & 0.116(1) \\
66.5     & 18.4(4)  & 3.0(7)  & 0.057(1) \\
\end{tabular}
\end{ruledtabular}
\end{table}

As seen from Table~\ref{table:VRH model}, the scaling temperature $T_0$ in this model does not appear independently of the power-law sensitivity $\nu$.  Instead, it appears only in the combined expression ${T_0}^{\nu}$. As a result, we do not assign theoretical significance to the particular numerical values found here for $T_0$ and $\nu$.   
For our purposes, it suffices to see that they combine to produce an overall decreasing logarithmic sensitivity $S$ with increasing metal percentage, as expected, and that the model given by Eqs.~(\ref{eq:Mobius}) and (\ref{eq:sensitivity}) give us a useful means of estimating the temperature dependence of $\rho(T)$ and $S$ from the measured room temperature resistance.  

The values for $\rho_0$ 
of Table~\ref{table:VRH model} decrease as the metal percentage increase, as expected. However, the listed $\rho_0$ values are still an order of magnitude larger than the resistivity $\rho \approx 1.1\,\mu \Omega$m for nichrome (Ni$_{0.8}$Cr$_{0.2}$) \cite{Mooij1973a}. From the TEM images, we see that the cermet structure consists of metal grains wrapped by insulator (rather than a uniform mixture of metal and insulator grains). Further, the 0.4 nm mean free path in NiCr \cite{Mooij1973a} is much smaller than  the 2 to 5 nm NiCr metal grain size, indicating that the enhanced resisitivity is an intergrain effect \cite{RevModPhys.79.469}. We thus tentatively attribute the enhancement of ${\rho}_0$ to the proximity to the percolation threshold with narrow conduction paths through the material \cite{McAlister1985}. 

\begin{figure}[tb!]
\includegraphics[width = 0.45
\textwidth]{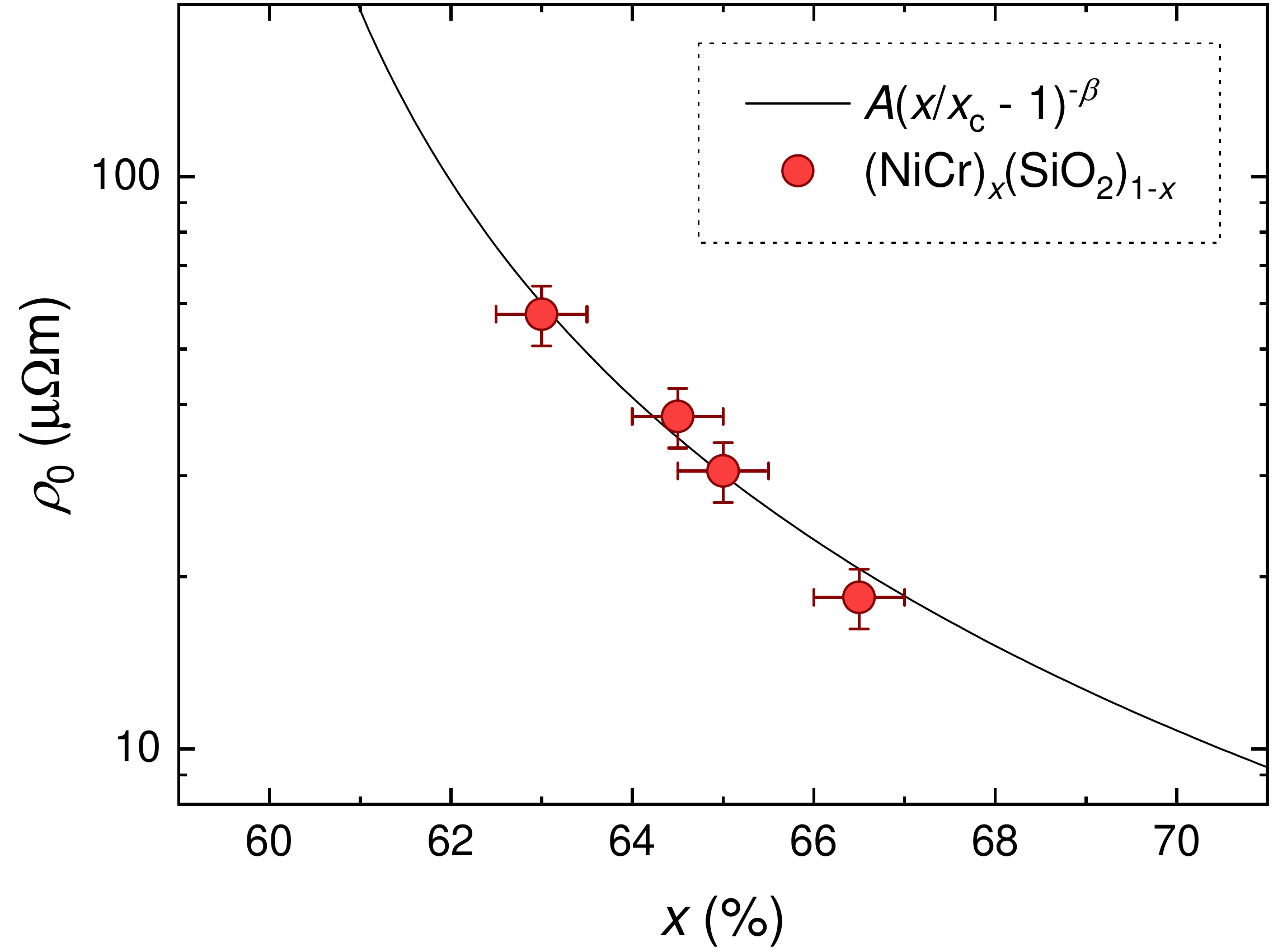}
\caption{Fit of percolation model Eq.~(\ref{eq:percolation}) for dependence of residual resistivity $\rho_0$ to data presented in Table~\ref{table:VRH model}.  The calculated curve corresponds to a percolation threshold $x_c = 0.59\pm0.1$ with $\rho_0$ set equal to the bulk resistivity for NiCr in the $x = 1$ continuous thin film limit.} 
\label{fig:percolation_fit}
\end{figure}

In the transition region of percolation near the percolation threshold \cite{McAlister1985, Dorfman1998, LEWIS2009215},  the conduction path contribution to resistivity
\begin{equation}
    \label{eq:percolation}
    \rho(x, T) \bigg \vert_{x \rightarrow x_c}  =  \alpha(T) \left(x/x_c - 1 \right)^{-\beta} \quad\mathrm{ for\ } x > x_c
\end{equation}
where  $x$ is the metal percentage,  $x_c$ is the percolation threshold ---  the metal percentage at which a narrow but continuous transport network is formed by the metallic grains --- and $\alpha(T)$ represents the temperature dependence for a given $x$ and $x_c$.  $\beta$ is a critical exponent that depends on dimensionality, with $\beta \rightarrow 2$ in the 3D limit \cite{McAlister1985, LEWIS2009215}. 

As the metal percentage decreases  as $x$ approaches $x_c$, the resistivity $\rho(x)$ at any fixed $T$ increases, and thus so does $\rho_0$ from our model. The critical value $x_c$ depends on dimensionality and sample preparation conditions, including deposition and annealing temperatures, as these can affect cluster sizes and distributions. Experimentally \cite{McAlister1985}, values ranging from 0.54 to 0.62 were observed for co-sputtered (Au, SiO$_2$). Here, if our materials are in this limit, then $x_c \le 62.5\%$.  A representative fit of Eq.~(\ref{eq:percolation}) to the data in Table~\ref{table:VRH model} with $x_c = 0.59\pm0.1$, $\beta = 1.7\pm0.2$  and $\rho_0$ set equal to  $1.15\,\mu\Omega\cdot m$ in the continuous NiCr thick film $x = 1$ limit is shown in Fig.~\ref{fig:percolation_fit}.

In percolation theory \cite{McAlister1985}, there is an intermediate weak-localization region between the classical percolation limit $x_c$ and classical metallic behavior for $x \ge x_{\mathrm{MIT}}$. As a result, numerical value for the percolation threshold $x_c$ calculated from the inferred high temperature residual resistivity $\rho_0$ using Eq.~(\ref{eq:Mobius}) will be less than the numerical value for the metal-insulator transition $x_{\mathrm{MIT}}$ inferred from the measured low-temperature logarithmic sensitivity. Assigning $x_{\mathrm{MIT}}$ to the metal concentration at which $dS/dT$ changes sign from positive to negative \cite{mobius_metal-insulator_2018}, we find $x_{\mathrm{MIT}} \approx 0.665$. 
 Thus, with the exception of the $x = 0.68$ film, the (NiCr, SiO$_2$) cermets presented here  are in that intermediate regime  $x_c < x < x_{\mathrm{MIT}}$.

\subsection*{Cermets sputtered in a mixed argon - oxygen atmosphere}

\begin{figure}[t!]
\includegraphics[width = 0.45\textwidth]{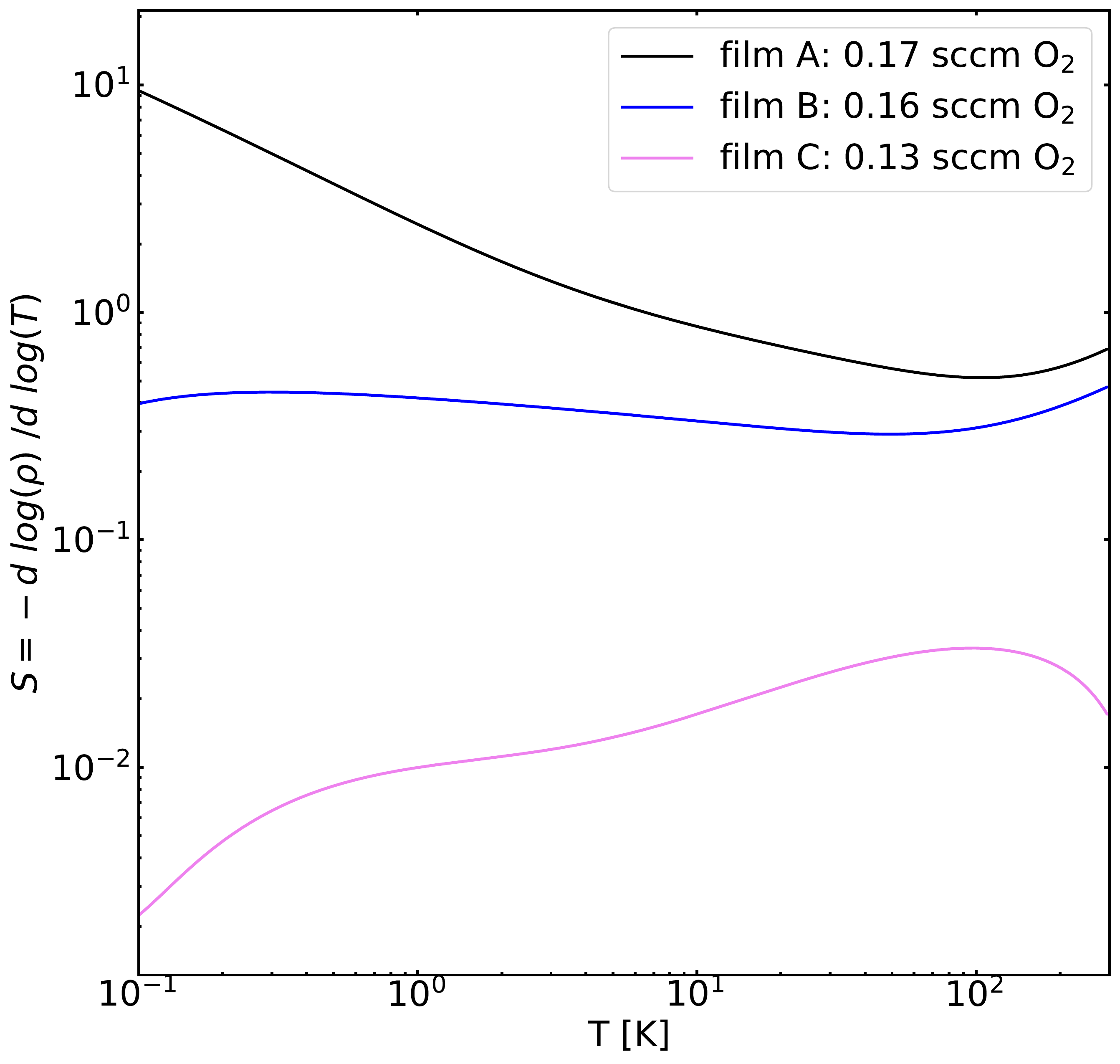}
\caption{Logarithmic (dimensionless) sensitivity $S$ for a series of films sputtered in a mixed argon + oxygen atmosphere at differing oxygen flow rates. These films correspond to the $\rho(T)$ graphs for films A, B, and C shown in Fig.~2 of the main text.   The slope $dS/dT$ changes from negative to positive   as the films change from metallic to insulating with increasing oxygen flow rate. The increase in sensitivity at high temperature that occurs in films A and B is not accounted for by the extended variable range hopping model in Fig.~\ref{fig:Mobiusfit}.}
\label{fig:S_for_oxygen_samples}
\end{figure}

The effect of co-sputtering in a  reactive oxygen + argon atmosphere on the temperature dependence of $\rho(T)$ is discussed in detail in the main text. Here we add a graph of the dependence of logarithmic sensitivity $S$ for the insulating-side films shown Fig.~2 of the main text. As shown in Fig.~\ref{fig:S_for_oxygen_samples}, the sensitivity increases as oxygen concentration increases. Attractively, film B exhibits a nearly constant sensitivity $S \approx 0.4$ between 0.1\,K and 100\,K.  The increase in sensitivity with increasing temperature above $100$\,K for the insulating side films indicates that sputtering in a reactive oxygen atmosphere introduces new physics near room temperature not included in Eq.~(\ref{eq:Mobius}).
\newpage
\begin{acknowledgments}
 
Support from the Knut \& Alice Wallenberg Foundation under Grant No.\,KAW\,2018.0019 (N.K., A.R.) and the Swedish Research Council, D.Nr.\,2021-04360 (A.R.), is acknowledged. A portion of this work was performed at the National High Magnetic Field Laboratory, which is supported by National Science Foundation Cooperative Agreements No. DMR-1644779 and DMR-2128556 and the State of Florida.  This work was also supported by the NSF MRI program, award number 2018560. N.F., J.P.-F., and N.K. contributed equally to this work.

\end{acknowledgments}

\bibliography{cermetpaper_local_full}

\end{document}